# Tailoring Photocatalytic Water Splitting Activity of Boron Thiophene Polymer through Pore Size Engineering

Preeti Bhauriyal,[a] Thomas Heine*,[a, b, c, d]

[a] Faculty of Chemistry and Food Chemistry, Technische Universität Dresden, Bergstrasse 66, 01069 Dresden, Germany

[b] Helmholtz-Zentrum Dresden-Rossendorf, HZDR, Bautzner Landstr. 400, 01328 Dresden, Germany

[c] Center for Advanced Systems Understanding, CASUS, Untermarkt 20, 02826 Görlitz, Germany

[d] Department of Chemistry and ibs for nanomedicine, Yonsei University, Seodaemun-gu, Seoul 120-749, South Korea

**E-mail:** thomas.heine@tu-dresden.de

**Abstract**

Taking into account the electron-rich and visible light response of thiophene, first-principles calculations have been carried out to explore the photocatalytic activity of donor-acceptor polymers incorporating thiophene and boron. The designed honeycomb-kagome boron thiophene (BTP) polymers with varying numbers of thiophene units and fixed B center atoms are direct band gap semiconductors with tunable band gaps ranging from 2.41 eV to 1.88 eV, and show high absorption coefficients under the ultraviolet and visible regions of the solar spectrum. Fine-tuning the band edges of the BTP polymer is efficiently achieved by adjusting the pore size through the manipulation of thiophene units between the B centers. This manipulation, achieved without excessive chemical functionalization, facilitates the generation of an appropriate quantity of photoexcited electrons and/or holes to straddle the redox potential of the water. Our study demonstrates that two units between B centers of thiophene in BTP polymers enable overall photocatalytic water splitting, whereas BTP polymers with larger pores solely promote photocatalytic hydrogen reduction. Moreover, the thermodynamics of hydrogen and oxygen reduction reactions proceed either spontaneously or need small additional external biases. Our findings provide the rationale for designing metal-free and single-material polymer photocatalysts based on thiophene, specifically for achieving efficient overall water splitting.

## 1. Introduction

Photocatalytic water-splitting for solar-driven hydrogen generation is a prospective and sustainable technology for clean and sustainable generation of chemical energy. Since the very first report of $TiO_2$ as a photocatalyst,[1] the progress of technically and economically applicable photocatalytic technology has become a top priority in sustainable energy research. It crucially relies on the development of photocatalytic materials offering low overpotential due to appropriate electronic structure and optimized surface reactivity, high thermal and chemical stability, high target selectivity to mitigate side reactions, and economically viable materials. Recently, organic semiconductors have emerged as promising materials for photocatalytic hydrogen and oxygen evolution, because their advantages include the absence of metals, high porosity, flexible geometrical topologies, a full spectrum of UV-vis activity, and tunable electronic properties.[2-5] Poly(p-phenylene) was first reported as a photocatalyst for hydrogen evolution in the year 1985, but its activity was poor and limited to the ultraviolet spectrum.[6-7] The potential for superior molecular design has brought the breakthrough in π-conjugated metal-free catalysts with the easy modulation of their photophysical and photochemical properties. Therefore, in recent years various metal-free polymers such as carbon nitrides,[8-9] conjugated microporous polymers (CMPs),[10-11] linear conjugated polymers[12-13] and covalent triazine-based frameworks (CTFs)[14-15] have been designed and studied in photocatalysis. Recently, thiophene (TP) has been widely applied as a key structural element for opto-related electronic polymers, serving as the strong electron-donating chromophoric center to collect photons. In this aspect, diverse organic polymers that contain TP units have been synthesized through C–C coupling, C–H activation, cyclotrimerization of the cyano group, and imine linkages. The incorporation of thiophene groups into photocatalytic materials adapts the electronic structure, tuning LUMO and HOMO positions, and narrows the band gap. The enhanced photocatalytic performance of materials modified with thiophene groups compared to other photocatalysts has been demonstrated recently. For example, instructive works by Liu's group have explored thiophene-based conjugated porous polymers (CPPs) for photocatalytic hydrogen evolution under visible light irradiation with excellent acceleration by tuning the composition of water-based co-solvents.[16-17] Amongst the various related works,[18-20] Chou et. al. observed the best HER performance (97.1 mmol/hg) in sulfone-based 4,8-bis(5-(2-ethylhexyl)thiophen-2- yl)benzo[1,2-b:4,5-b']dithiophene (BDTT) conjugated polymer under visible-light illumination and an apparent quantum yield exceeding 18% at a wavelength of 500 nm.[20] However, most TP polymers only have band edges covering water reduction potentials to trigger a HER half-reaction of photocatalytic water splitting and have insufficient drive force provided by the irradiation-induced holes. To realize photocatalytic overall water splitting under visible light with a metal-free photocatalyst, it is usually required to build heterostructures with two catalysts, one for HER and one for OER. Very recently, such heterojunctions have also been realized in a two-dimensional (2D) single-material photocatalyst. Two 2D covalent organic frameworks (COFs), namely 1,3,5-triethynylbenzene polymer (PTEB) and 1,3,5-tris-(4-ethynylphenyl)-benzene polymer (PTEPB), are reported to split water under visible light without using any sacrificial reagents and cocatalyst.[21] Jing et. al. proposed a high quantum efficiency overall photocatalyst by coupling B- and N-HTs into a honeycomb-kagome structure with alternating distribution of heteroatoms.[22] A similar strategy can be applied to the TP polymers by introducing B at alternate positions. In these polymers, the boron atoms can function as Lewis acid sites, augmenting catalytic activity. The experimental support for such a combination has been validated by the research of Ren and Jäkle groups, which have a long-standing interest in incorporating boron into oligo- and polythiophenes to construct new functional materials that can harvest the unique chemical and electronic characteristics of organoboranes.[23] Interestingly, Ren et al. recently suggested a new synthetic strategy to incorporate a boron-containing building block into conjugated porous thiophene polymers by using efficient boron/tin (B/Sn) exchange reaction.[24] The conjugated structure in these polymers facilitates electron transfer, promoting light absorption, and making them attractive candidates for photocatalytic applications.

In this work, we theoretically explore the chemistry of such B-TP polymers, and demonstrate the remarkable ability of size variation coupled with an alternating arrangement of donor-acceptor units (TP and B) to tune the photocatalytic activity for water splitting. The study involves the detailed structural evaluation of BTP polymers with varying numbers of TP units in a kagome lattice with a honeycomb sublattice that carries the B center atoms. The analysis of the electronic structure reveals that by simply adjusting the coupling between the donor and acceptor, it is possible to achieve the desired alignment of the band edges with respect to the targeted electrode potentials for the HER and OER processes, absorption spectra, and solar-to-hydrogen (STH) efficiency for overall photocatalytic water splitting. Finally, for assessing the solar-driven water splitting efficiency, the thermodynamics of water oxidation and hydrogen reduction half-reaction mechanisms are investigated.

**2. Computational Details**

All calculations in this work are performed using Vienna ab initio Simulation Package (VASP) code based on density functional theory (DFT) formulated using periodic boundary conditions.[25-26] The generalized gradient approximation (GGA) involving the Perdew−Burke−Ernzerhof (PBE) functional with Grimme's D3 van der Waals correction was adopted to process the exchange−correlation term and accurately account for the long-range van der Waals (vdW) forces between the adsorbents and the BTP polymers.[27-28] The ions are modelled with the projector augmented-wave (PAW) method.[29-30] The plane wave cutoff energy was set to 500 eV, with the convergence threshold of force less than 0.001 eV/Å during geometry optimization. The Monkhorst-Pack k-point meshes are set at 7×7×1, 5×5×1, and 3×3×1 for BTP-1, BTP-2, and BTP-3, respectively. A vacuum space of 15 Å along the z-direction is adopted to avoid interactions between two layers in nearest-neighbouring unit cells. Self-consistent field calculations are performed with a convergence criterion of $1 \times 10^{-5}$ eV per atom. The Heyd−Scuseria−Ernzerhof screened hybrid functional (HSE06)[31] was employed for band structure predictions as the GGA-PBE functional usually underestimates the band gaps. In order to assess the ability of photocatalytic water splitting of 2D BTP polymers, the band edges were corrected by the vacuum level and compared with the water redox potentials. The free energy change (ΔG) during HER and OER on the surface of 2D nitrogen-linked COFs is calculated based on the computational hydrogen electrode (CHE) model developed by Nørskov et al.[32] Detailed computational methods are described in the Supporting Information. To explore the optical properties of the determined structure, the optical absorption coefficient as a function of photon energy is calculated with the HSE06 functional according to the following equation,[33]

$$\alpha(\omega) = \frac{\sqrt{2}\omega}{c}\left[\sqrt{\varepsilon_1^2 + \varepsilon_2^2} - \varepsilon_1\right]^{1/2} \quad (1)$$

where $\varepsilon_1$ and $\varepsilon_2$ are the real and imaginary parts of the dielectric function, respectively, ω is the photon frequency, and c is the speed of light, respectively. The polarization vector parallel to the z-axis is considered, so the $\varepsilon_1$ and $\varepsilon_2$ are averaged over two polarization vectors (along x and y directions).

## 3. Results and Discussions

### 3.1. Designing of Boron-Thiophene Polymers

Here, we have designed three B-TP polymers, BTP-1, BTP-2, and BTP-3, where the two neighboring B atoms are connected by mono-, bi-, and terthiophene units, respectively, as shown in Figure 1(a, d, and e). The strategy is in consequence of the study of Ren et al., which involves the synthesis of B-TP polymers of various pore sizes by tuning the TP-Sn building blocks, where the reaction condition of temperature and catalyst determine the structural arrangement of B and TP.[24] The rotation of the thiophene ring around the B-C bond can lead to symmetric or asymmetric arrangements of thiophene rings, resulting in different conformers, both of them with twisted thiophene rings. Our study focuses on the energetically more stable symmetric conformer, where the B and three neighboring C atoms are coplanar and the B atom is $sp^2$ hybridized (more details in Figure S1, Supporting Information). This symmetric arrangement has been synthesized by strategically optimizing reaction conditions (dichlorobenzene and 180 °C temperature).[24] In energetically stable symmetric BTP-1 polymer, all S atoms in nonplanar thiophenes are aligned in one direction (Figure 1(a)) with a single layer thickness of 1.59 Å, however, with no unsaturated dipoles as observed in plane-averaged electrostatic potential plot illustrated in Figure 1(b). The electron localization function (ELF) is used to determine the charge distribution and chemical bonding character (Figure 1c). Bonding electron pairs are separated from C to S because of their different electronegativity, indicating the formation of polar covalent bonds. The C-C and B-C bonds in BTP polymer are the typical covalent ones. Regarding the BTP-2 and BTP-3 polymers, the rotation of the TP unit can generate different structural conformers (Figure S2-S4, Supporting Information). Specifically, we have investigated the conformers (trans conformers depicted in Figure 1(d and e)) that exhibit the highest energetic stability. The stability of BTP polymers is further explored through molecular dynamics simulations, conducted at 300 K over a 20 ps duration (Figure S5). Throughout the simulation, there is a subtle twisting of TP rings, but the deviation from the initial local minima is minimal (simulation trajectories are available, see data availability statement section). Notably, the transition state for conformational change (Figure S1) remains elusive, possibly attributed to a higher energy barrier (0.21 eV) than the thermal energy at room temperature (0.025 eV). Overall, there are no geometric reconstructions, which reveals the good thermal stability of monolayer of BTP polymers for room temperature applications.

All three BTP polymers have a kagome lattice (red lattice in Figure 1d) with a honeycomb sublattice that carries the B center atoms. The optimized lattice parameters are found to be 9.60 Å, 16.32 Å and 23.17 Å for BTP-1, BTP-2 and BTP-3, with respective pore sizes of 5.64 Å, 12.09 Å, and 18.36 Å (more structural details revealed in Table S1).

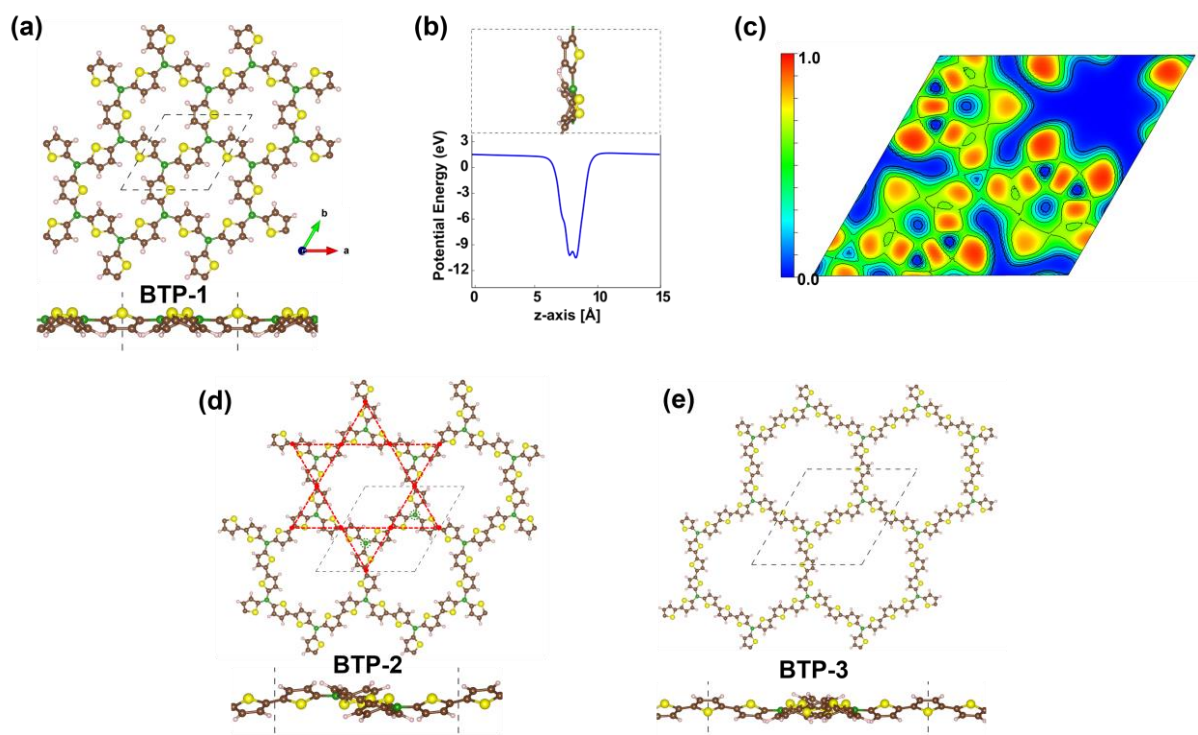

**Figure 1:** (a) Top and side views of BTP-1 polymer. (b) Plane-averaged electrostatic potential plot as a function of the distance in z-direction and (c) electron localization function (ELF) plot of BTP-1 polymer. Top and side views of (d) BTP-2 and (e) BTP-3 polymers.

### 3.2. Electronic and Optical Properties

The foundation of photocatalysts is typically based on the broad absorption of light to generate photon-induced excitons and the band alignment with respect to the $H_2O$ oxidation and reduction potentials. Considering the direct relation of absorption capability with the band gap, we first carried out a detailed investigation of band structures along the symmetry line in the Brillouin zone, from Γ to M and K, and density of states (DOS), calculated at HSE06 level of theory. The obtained band structures (Figure 2) show that BTP-1, BTP-2, and BTP-3 are direct band semiconductors with 2.41, 2.07, and 1.88 eV band gaps, respectively, where both conduction band minima (CBM) and valence band maxima (VBM) (if they can be specified) lie at Γ point. The band structures of BTP polymers show characteristics of a kagome lattice in that it exhibits Dirac cones that lie below flat bands in both the conduction and valence band regions. The reason for this is that the TP bridges of BTP polymers define a kagome lattice (Figure 1). On a more closer look, the honeycomb sublattices (Figure S6) occupied by the empty π-orbitals of B, selectively contribute to the conduction band maximum around the Fermi level as can be seen from the partial density of states and charge density distribution analysis (Figure 2). These BTP polymers can be thus defined as honeycomb-kagome polymers.[34] In addition to lattice symmetry,[35-37] the frontier molecular orbital (MO) symmetries of the building monomers are also expected to be a key factor determining the nature of the electronic bands near the Fermi level.[38, 39] The analysis of molecular fragment (Figure S7) shows that the kagome band is originated from the single (non-degenerate) lowest unoccupied molecular orbital (LUMO) level plus doubly degenerate LUMO+1 level. A doubly degenerate highest occupied molecular orbital (HOMO) level in the molecular unit leads to a flat highest VB with two Dirac bands below it.[38] The resultant VBMs are degenerate at the $\Gamma$ point with the flat bands having hole effective masses from 19.0, 36.1, 110.8 $m_0$, and dispersive sub-bands with 0.37, 0.97, 0.45 $m_0$ for BTP-1, BTP-2 and BTP-3, respectively. Both light and heavy mass carriers could therefore coexist at the $\Gamma$ point. The flat valence bands show no significant influence of the B atoms but are delocalized on the remainder of the 2D lattice. The pronouncedly dispersed CBM are associated with effective $\pi$-conjugation in the 2D lattices and low electron effective masses, covering the range of 0.28 – 0.31 $m_0$ (Table 1).

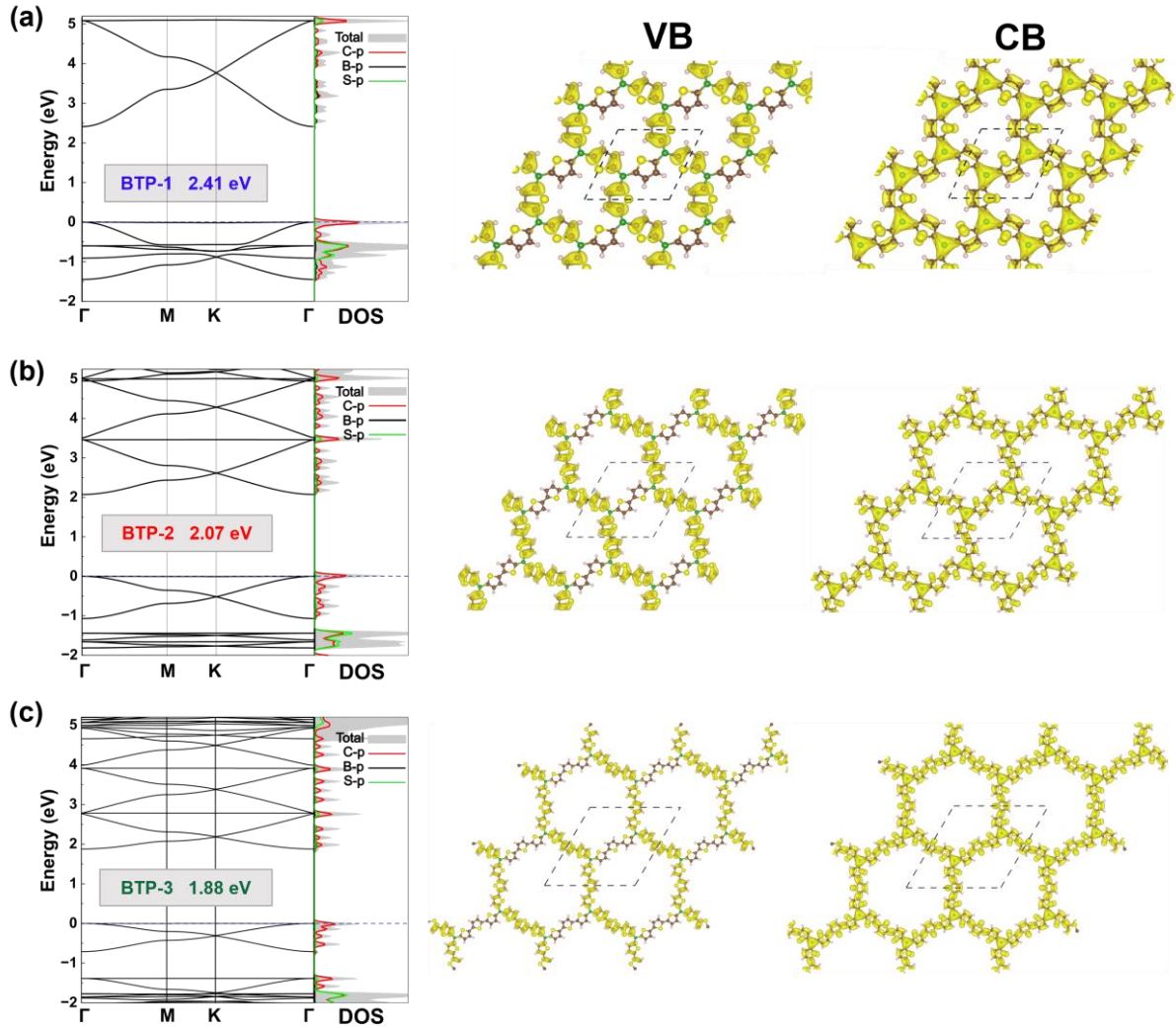

**Figure 2:** Band structures, at the HSE06 level, and corresponding charge density distribution for the CBM and VBM, respectively, of honeycomb-kagome BTP polymers

**Table 1:** Calculated band gaps, effective masses and drive potential of photogenerated electrons ($U_{red}$) and holes ($U_{ox}$).

| 2D Polymers | Band Gaps (eV) | $m^*$ ($m_0$) | $U_{red}$ (eV) | $U_{ox}$ (eV) |
|---|---|---|---|---|
| **BTP-1** | 2.41 | 0.27 ($m_e$)<br>18.99 ($m_h$) | 0.10 | 2.32 |
| **BTP-2** | 2.07 | 0.29 ($m_e$)<br>0.97 ($m_h$) | 0.88 | 1.30 |
| **BTP-3** | 1.88 | 0.31 ($m_e$)<br>0.45 ($m_h$) | 1.02 | - |

It is well known that for materials to be good photocatalysts, the band gap should be moderate ~1.5 - 3.0 eV, and band edge positions of the VBM and CBM should exceed the water redox potential. For BTP polymers, band edge alignments with respect to water reduction and oxidation potentials at pH = 0 (-4.44 eV, -5.67 eV) are illustrated in Figure 3a. The modulation of pore size based on thiophene rings not only affects the band gap values but also drastically tunes the band edges. We find that the energies of both CBM and VBM edges increase from BTP-1 to BTP-3 polymers. This shift is associated with the alteration of the chemical potential of the HOMO at the molecular level (Figure S8). The rise in the number of electron-donor thiophene rings from BTP-1 to BTP-3 selectively elevates the HOMO energy level without affecting the LUMO, ultimately resulting in a reduction of the band gap. At pH=0, CBM and VBM of BTP-1 and BTP-2 can straddle the full redox potential to be used as a comprehensive water-splitting photocatalyst. On the other hand, BTP-3 has the potential to drive only the HER process. In the process of photocatalytic water splitting, the potential energy provided by photogenerated carriers directly decides whether the half-reactions for water splitting can occur spontaneously. The energy difference between the CBM and the potential of $H_2/H_2O$ determines the potentials of photogenerated electrons for HER ($U_e$), while the energy difference between the VBM and the potential of $H_2/H_2O$ determines the potentials of photogenerated holes for OER ($U_h$) (shown in Figure 3a). The calculated values of $U_e$ and $U_h$ for BTP polymers at pH 0 are given in Table 1.

A high-efficiency photocatalyst should have a strong ability to capture sunlight, especially in the visible and ultraviolet regions. Thus, we have further illustrated its optical response performance by calculating the absorption spectra as shown in Figure 3b. All three polymers show strong absorption in the ultraviolet and are visible with an absorption coefficient of ~$10^5$ cm$^{-1}$, and the pore size increase exhibits a red shift in the absorption peaks of BTP polymer. The absorption is much higher than in g-$C_3N_4$[40] and comparable to that of organic perovskite solar cells.[41] The STH conversion efficiency of all three BTP polymers is calculated using the following equation:

$$STH = \frac{\int_0^{edge} 1.23 \times I(\lambda)\lambda d\lambda}{\int_0^{2000} 1240 \times I(\lambda) d\lambda} \times QE \qquad (2)$$

where, λ represents the wavelength of light. The value of 2000 nm is selected as the maximum wavelength within the solar spectrum. QE denotes the quantum efficiency, which is assumed to be 100%. I(λ) represents the intensity of blackbody radiation at 6,000 K, which is used to simulate the solar spectrum. The theoretically calculated STH of BTP-1, BTP-2, and BTP-3 are calculated to be catalyst can reach 10.6%, 15.8% and 20.2%, respectively.

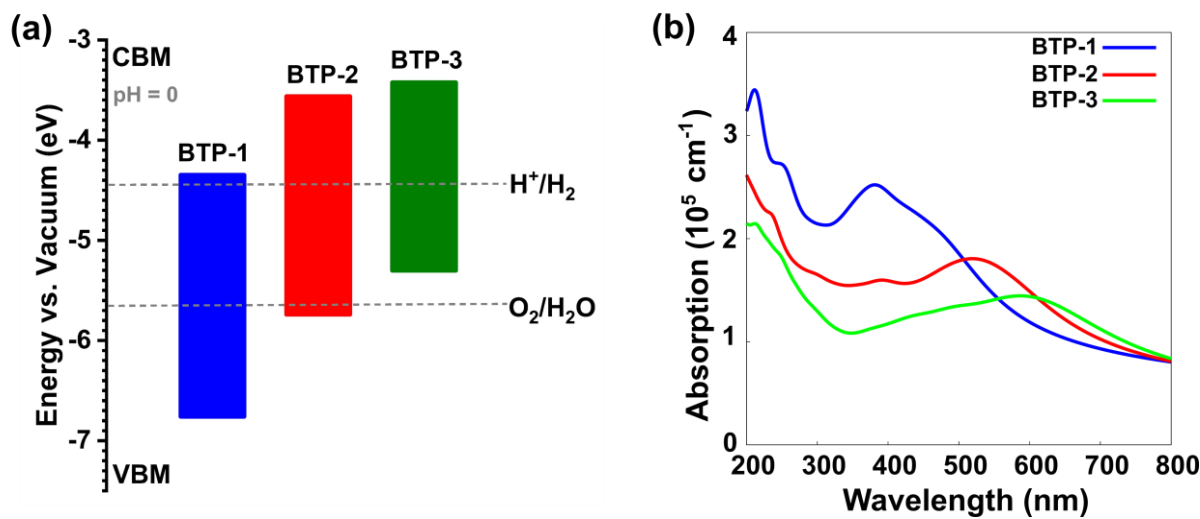

**Figure 3: (a)** Band alignments of 2D BTP polymers. The black dashed lines denote the redox potential of water at pH = 0. **(b)** The computed optical adsorption spectra of 2D BTP polymers for light-induced from the in-plane directions.

### 3.3. Thermodynamics of Hydrogen and Oxygen Evolution Reactions

Along with the appropriate band alignment, the ease of redox catalysis plays an important role in photocatalysis. This assessment can be conducted by calculating the thermodynamics of the photoelectrode processes using the computational hydrogen electrode model.[32] Here, the thermodynamics of the intermediate states of HER and OER processes is first calculated in the absence of light irradiation (U=0), followed by the presence of light at pH 0. Figure 4 shows the involved reaction mechanism and Gibbs free energy change for the elementary steps. The calculation details are included in the Supporting Information. The HER involves one intermediate state of *H absorbed on the BTP polymers as a two-electron reaction process. We have investigated all conceivable configurations for H* adsorption. Through a comparison of the adsorption energy at various potential active sites, it has been determined that the preferred adsorption site for the H* species is on the C edge of TP (Figure 4a). Without light-induced bias potential (U = 0), the free energy changes are 0.48, 0.51 and 0.57 eV, leading to unfavourable HER process. The photogenerated electrons provide sufficient energy ($U_e$) of

0.88 eV and 1.02 eV to reduce the free energy requirement for the elementary steps and the HER becomes completely spontaneous for BTP-2 and BTP-3. Although, for BTP-1, additional external bias of 0.38 V should be added to drive HER, which is comparable to Ni/graphene composite (0.35 V)[42] and g-$C_3N4$ (0.43 V)[43].

The OER is considered a four-electron reaction process, with two possible mechanisms single site (involving *OOH formation) or dual-site (with *O*OH formation) process, where there are several intermediate states of *OH, *O, *OOH, or *O*OH, and *O*O absorbed on the BTP polymers as shown in Figure 4a (details in Supporting Information). Since the 2D BTP-3 lacks suitable band edges for OER photocatalysis, our focus in this context shifts exclusively to the OER process occurring on BTP-1 and BTP-2 polymers. The Gibbs free energies for each intermediate state in Figure 4c and Table S3 reveal the feasibility of each elementary step. The first two steps are the dissociation of $H_2O$ with the formation of the adsorbed OH (OH*) on the active B site, and then the formation of the adsorbed O (O*) is accomplished by oxidizing OH*. It is found that O* adsorption is possible at both B-top and B-C bridge active sites, with the B-C bridge site being energetically more favourable. Even the energy barriers for the formation of O* using the nudged elastic band (NEB) method show the favourability (by an energy barrier of 2.75 eV) of the B-C bridge site (Figure S9, Supporting Information). In the third step, the OER exclusively follows the dual-site process (formation of O*OH*), eliminating the possibility of the single-site process. This is due to the instability and subsequent breakdown of the optimized structure with OOH* on the B-C bridge site. In the dual-site process, the formation of O*OH* is favoured followed by formation O*O* and release of $O_2$ for all BTP polymers. For both BTP-1 and BTP-2, the conversion of O* to O*OH* is the rate-determining step, with the limiting potentials of 1.99 V and 2.02 V, respectively. The extra potential provided by the photogenerated hole in BTP-1 (2.32 V) is sufficient to overcome the barrier, whereas in the case of BTP-2, applying an external potential of 0.72 V or co-catalyst (e.g. Pt)[44,45] can trigger OER.

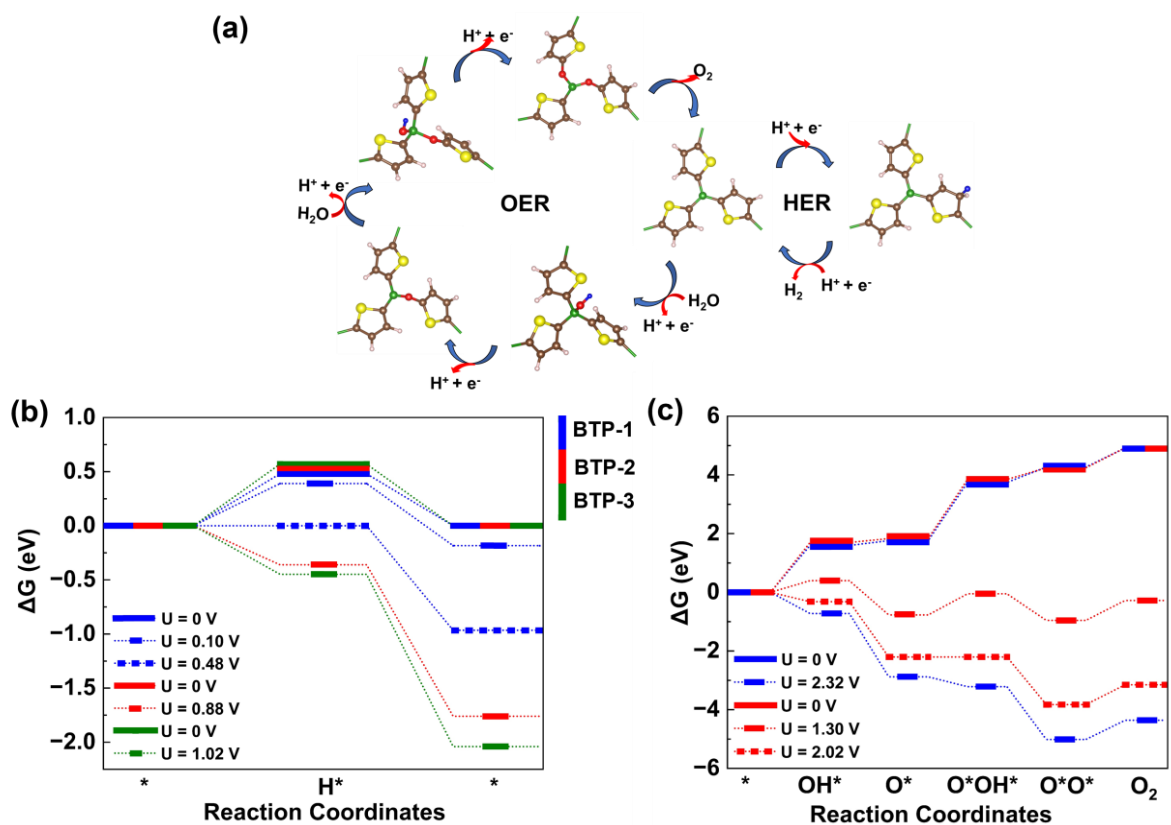

**Figure 4:** (a) Proposed photocatalytic pathway of water oxidation and hydrogen reduction of BTP polymers with the favourable atomic configurations of OH*, O*, O*OH*, O*O*, and H* intermediates. The adsorbed oxygen and hydrogen atoms are coloured red and blue, respectively. Calculated Gibbs free energy diagrams of (b) hydrogen reduction and (c) water oxidation of BTP polymers under the conditions of absence of light irradiation (U = 0 V), presence of light irradiation (U = $U_{ox}$ or $U_{red}$), and limiting potential.

These observations indicate that thiophene solely catalyzes the HER which is in agreement with the previous experimental studies,[16-20] however, the introduction of B facilitates the BTP polymer's ability towards OER by contriving donor-acceptor characteristics, also responsible for effective electron/hole separation. Both BTP-1 and BTP-2 are promising as overall water-splitting photocatalysts, however, as the pore size increase to BTP-3, only the photocatalytic HER is permitted.

## 4. Conclusion

In summary, we have systematically investigated the possible application of the new class of boron thiophene polymers to exploit their donor-acceptor character in the field of photocatalytic water splitting. These recently synthesized BTP polymers, namely BTP-1, BTP-2, and BTP-3, feature diverse pore sizes and exhibit a kagome lattice structure with a honeycomb sublattice containing B center atoms. These polymers display a distinctive kagome band structure characterized by Dirac cones situated beneath flat bands in both the conduction and valence band regions. All three BTP polymers are obtained to be direct band gap semiconductors with the range of 1.81-2.41 eV that shows strong optical absorbance in both ultraviolet and visible light regions. The CBM is quite dispersed offering mobile carrier ions with an effective mass of 0.27-0.31 $m_0$ compared to a rather flat VBM. The band alignment analysis shows that the increase in the donor thiophene units makes the polymer a better HER photocatalyst by destabilizing the CBM, thus increasing the external potential of photogenerated electrons, $U_{red}$ from 0.10 to 1.02 V for BTP-1 to BTP-3 (that makes HER spontaneous). The OER is catalyzed at the B center following a dual-site reaction mechanism, however, the VBM destabilizes with the increased pore size making BTP-3 unsuitable for photocatalytic OER. Overall, we identified that BTP-1 and BTP-2 can act as overall water-splitting photocatalysts and the BTP polymers having more TP units can only be a HER photocatalyst. Our observations suggest that experimentally realized BTP polymers can be used for photocatalytic water splitting. Moreover, by adjusting the pore size via controlled modification of thiophene units between B centers, without resorting to excessive chemical alterations, we can precisely fine-tune the band edges of the BTP polymer to straddle the redox potentials of the water.

**Data Availability Statement**

The geometries, corresponding band structures, and further calculations for all systems investigated in this work are available on zenodo repository as a dataset under doi:10.5281/zenodo.10467140.

**Supporting Information**

Isosurface representation of symmetric BTP molecule; rotation energy profiles for symmetric BTP to asymmetric BTP conversion; possible structural conformers and band structures of BTP polymers; AIMD simulation plot; BTP-2 polymer representing honeycomb kagome lattice; frontier orbitals of molecular fragment of BTP-1 polymer; frontier orbital energy level and band position for molecular fragments and corresponding BTP polymers; Reaction barriers form B-top *OH to B-top and bridge *O; Schematics of HER and OER on BTP-2 and BTP-3 polymers; Lattice parameters and pore size of BTP polymers; Free energy change of HER and OER for BTP polymers without light irradiation.

**Author Information**

**Thomas Heine** - Faculty of Chemistry and Food Chemistry, Technische Universität Dresden, Bergstrasse 66, 01069 Dresden, Germany; Helmholtz-Zentrum Dresden-Rossendorf, HZDR, Bautzner Landstr. 400, 01328 Dresden, Germany; Center for Advanced Systems Understanding, CASUS, Untermarkt 20, 02826 Görlitz, Germany; Department of Chemistry and ibs for nanomedicine, Yonsei University, Seodaemun-gu, Seoul 120-749, Republic of Korea; orcid.org/0000-0003-2379-6251.

Email: thomas.heine@tu-dresden.de

**Preeti Bhauriyal** - Faculty of Chemistry and Food Chemistry, Technische Universität Dresden, Bergstrasse 66c, 01069 Dresden, Germany; orcid.org/0000-0003-4767-6756.

E-mail: preeti.bhauriyal@tu-dresden.de

**Notes**

The authors declare no competing financial interests.

**Acknowledgment**

We thank the Center for Information Services and High-Performance Computing (ZIH) at TU Dresden and Noctua 2 at the NHR Center PC2 for the generous allocation of computer time. P. B. acknowledges the Alexander-von-Humboldt Foundation and Maria Reiche Postdoctoral Fellowships (granted by graduate academy TU Dresden) for funding. The authors also thank Deutsche Forschungsgemeinschaft within CRC 1415 and RTG 2861 for support.

# Supporting Information

# Tailoring Photocatalytic Water Splitting Activity of Boron Thiophene Polymer through Pore Size Engineering

Preeti Bhauriyal,[a] Thomas Heine*,[a, b, c, d]

[a] Faculty of Chemistry and Food Chemistry, Technische Universität Dresden, Bergstrasse 66, 01069 Dresden, Germany

[b] Helmholtz-Zentrum Dresden-Rossendorf, HZDR, Bautzner Landstr. 400, 01328 Dresden, Germany

[c] Center for Advanced Systems Understanding, CASUS, Untermarkt 20, 02826 Görlitz, Germany

[d] Department of Chemistry and ibs for nanomedicine, Yonsei University, Seodaemun-gu, Seoul 120-749, South Korea

**E-mail:** thomas.heine@tu-dresden.de

**Computational details for free energy change**

HER process could be decomposed into two one-electron steps with each step consuming a proton and an electron:

$$* + H^+ + e^- = H^* \quad (1)$$

$$H^* + H^+ + e^- = * + H_2 \quad (2)$$

OER process could be decomposed into four one-electron oxidation steps, corresponding to the deprotonation of water molecules, as follows:

$$* + H_2O = OH^* + H^+ + e^- \quad (3)$$

$$OH^* = O^* + H^+ + e^- \quad (4)$$

$$O^* + H_2O = OOH^* + H^+ + e^- \quad (5)$$

$$OOH^* = * + O_2 + H^+ + e^- \quad (6)$$

Dual-site

$$O^* + H_2O = O^*OH^* + H^+ + e^- \quad (7)$$

$$O^*OH^* = O^*O^* + H^+ + e^- \quad (8)$$

$$O^*O^* = * + O_2 \quad (9)$$

where, * denotes a site on the surface, *(radical) denotes the corresponding radical adsorbed on the surface.

we employed the computational hydrogen electrode model that proposed by Nørskov et al.[1] to compute the Gibbs free energy change of each elementary step of HER and OER.

To calculate the free energy changes involved in OER and HER process, Gibbs free energy change are calculated as,

$$\Delta G = \Delta E + \Delta E_{ZPE} + T\Delta S + \Delta G_{pH} + \Delta G_U \quad (10)$$

Here, $\Delta E$, $\Delta E_{ZPE}$ and $\Delta S$ is the total energy obtained from DFT calculations, zero-point energy and entropy at 298.15K. $\Delta G_{pH} = 0$ at pH = 0. $\Delta G_U$ refers to extra potential bias provided by an electron in the electrode, where U is the electrode potential relative to the standard hydrogen electrode (SHE). The light-induced driven potentials for HER, $U_{red}$ are defined as the energy differences between the hydrogen reduction potential and CBM. And, the potentials of photogenerated holes for the OER ($U_{ox}$) are defined as the energy differences between the VBM and hydrogen reduction potential.

Gibbs free energy changes, ΔG for each elementary step under the effect of pH and an extra potential are defined by the following equations:

$$\Delta G_1 = G(H^*) - G(*) - 1/2\, G(H_2) - eU_{red} \quad (11)$$

$$\Delta G_2 = G(*) - G(H^*) + 1/2\, G(H_2) - eU_{red} \quad (12)$$

$$\Delta G_3 = G(OH^*) + 1/2\, G(H_2) - G(*) - G(H_2O) - eU_{ox} \quad (13)$$

$$\Delta G_4 = 1/2\, G(H_2) + G(O^*) - G(OH^*) - eU_{ox} \quad (14)$$

$$\Delta G_6 = 1/2\, G(H_2) + G(O^*OH^*) - G(O^*) - G(H_2O) - eU_{ox} \quad (15)$$

$$\Delta G_7 = 1/2\, G(H_2) + G(O^*O^*) - G(O^*OH^*) - eU_{ox} \quad (16)$$

$$\Delta G_8 = G(O_2) + G(*) - G(O^*O^*) \quad (17)$$

For those reactions involving the release of protons and electrons, the free energy of one pair of proton and electron (H$^+$ + e$^-$) under standard conditions was taken as $1/2\, H_2$. The entropies of the free molecules, such as $H_2$, $H_2O$, were referenced to the NIST database. As the DFT method cannot accurately describe the high-spin ground state of the $O_2$ molecule, the Gibbs free energy of $O_2$ [$G(O_2)$] is obtained by $G(O_2) = 2G(H_2O) - 2G(H_2) - 4.92$.

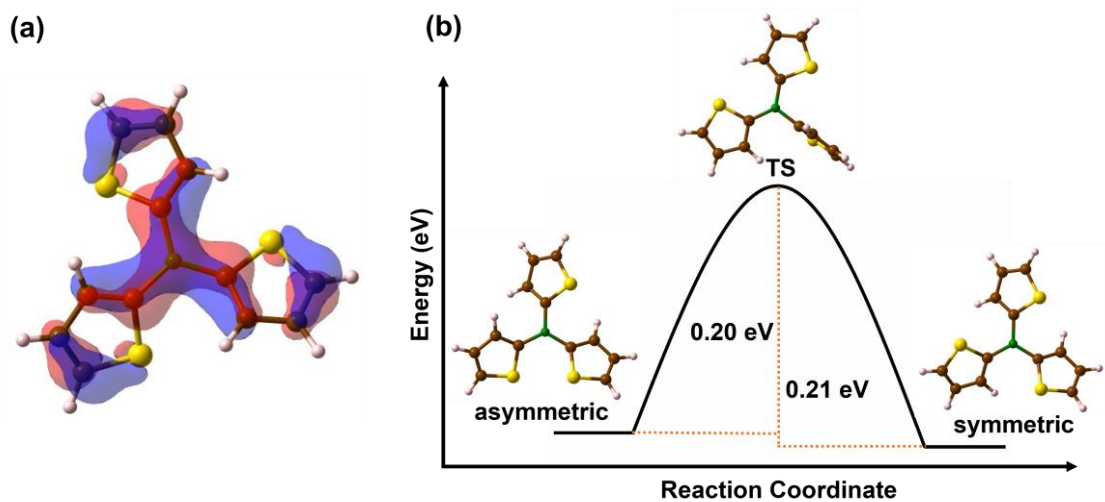

**Figure S1:** (a) Isosurface representation of the delocalized bonding orbitals ($p_z$) in symmetric BTP molecule and B is $sp^2$ hybridized. (b) Rotation energy profiles for symmetric BTP to asymmetric BTP conversion. Calculations are performed in Gaussian/g09.d01 at B3LYP/6-31+G* level theory.

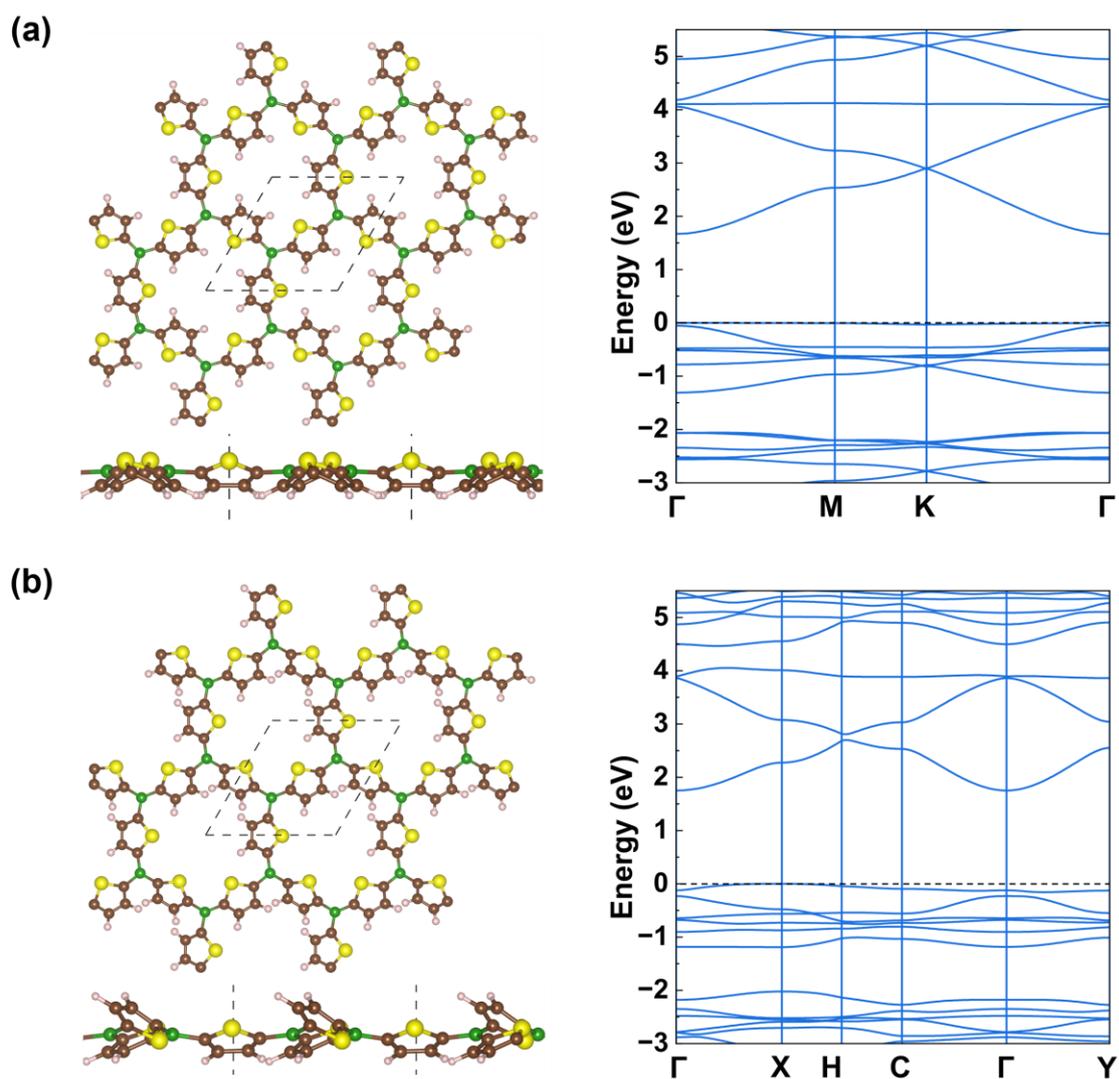

**Figure S2:** Two possible (a) symmetric and (b) asymmetric structural arrangements of BTP-1 polymer and the corresponding PBE band structures.

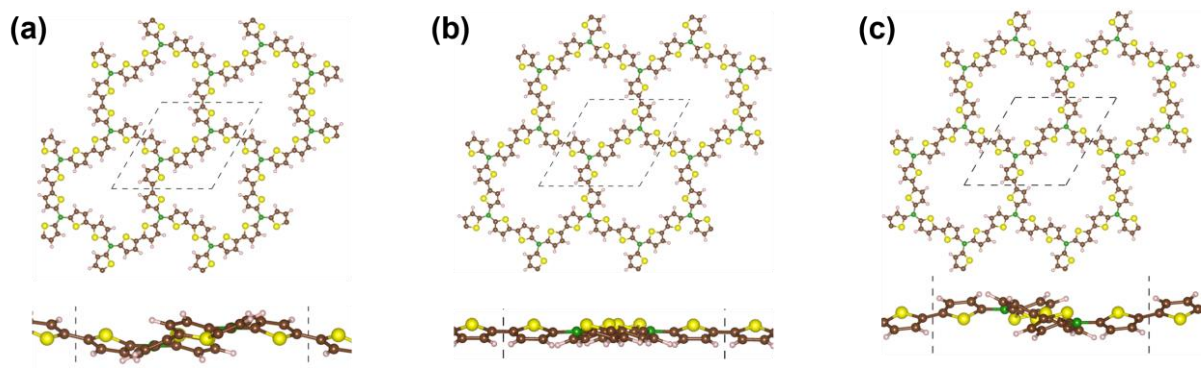

**Figure S3:** Three possible (a) asymmetric, (b) symmetric (alignment of S in one direction), and (c) symmetric (alignment of S in opposite direction) structural arrangements of BTP-2 polymer.

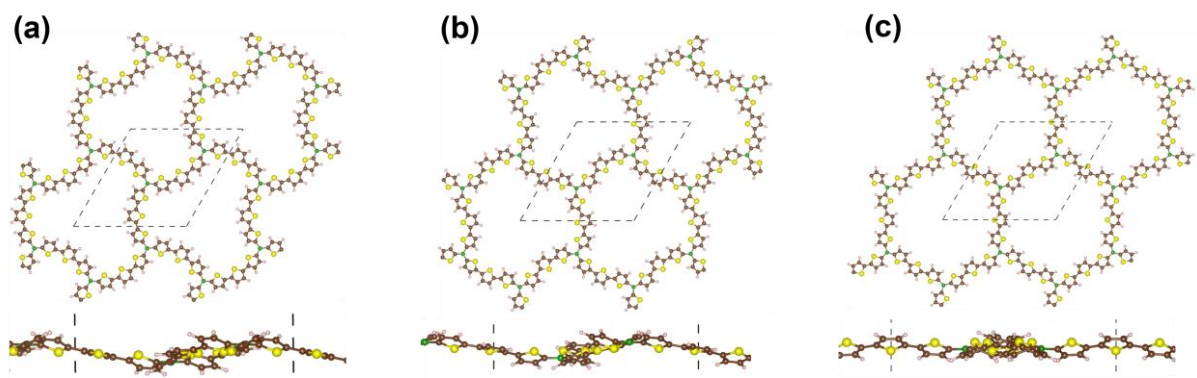

**Figure S4:** Three possible (a) asymmetric, (b) symmetric (alignment of S in one direction), and (c) symmetric (alignment of S in opposite direction) structural arrangements of BTP-3 polymer.

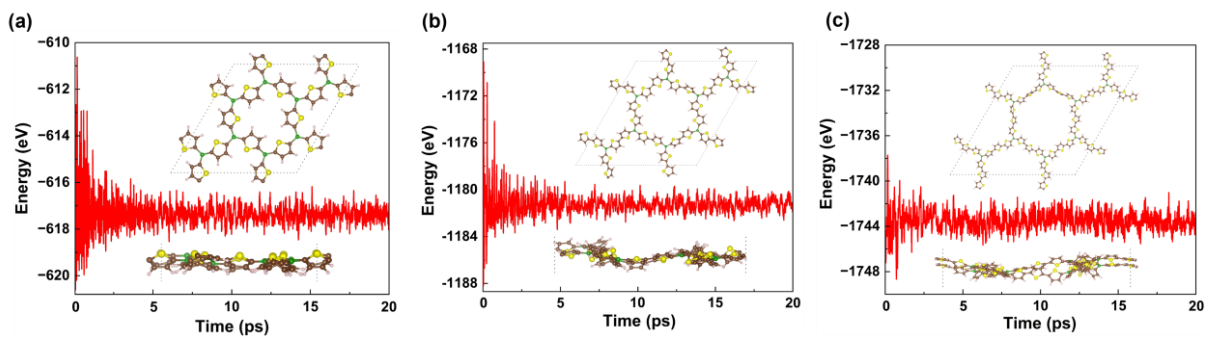

**Figure S5:** Total energy vs time variation during AIMD simulation of (a) BTP-1, (b) BTP-2, and (c) BTP-3, and their final snapshot frames at 300 K after 20 ps.

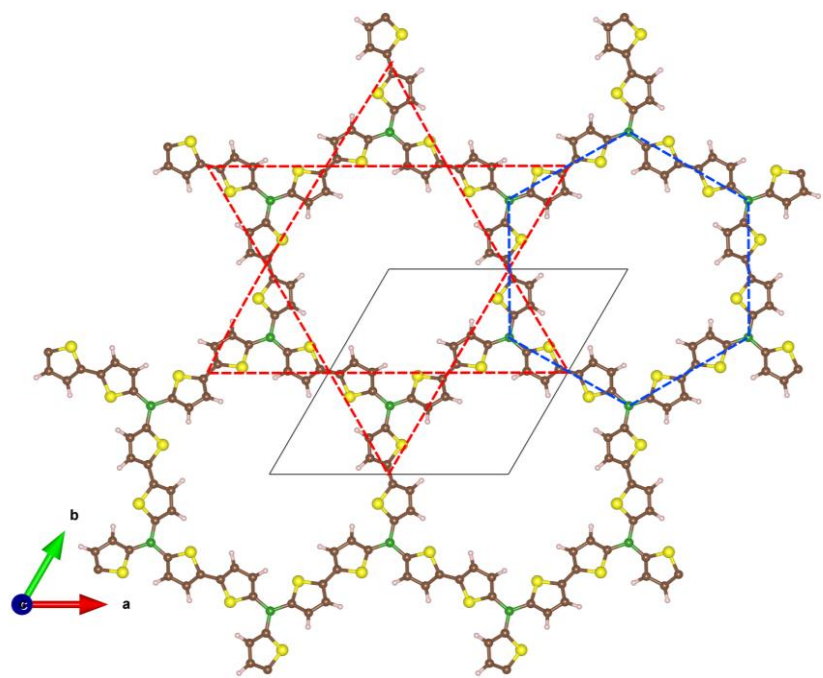

**Figure S6:** BTP-2 polymer representing honeycomb kagome lattice, a typical feature of BTP polymers. Kagome lattice and honeycomb sublattice are shown by red and blue colours, respectively.

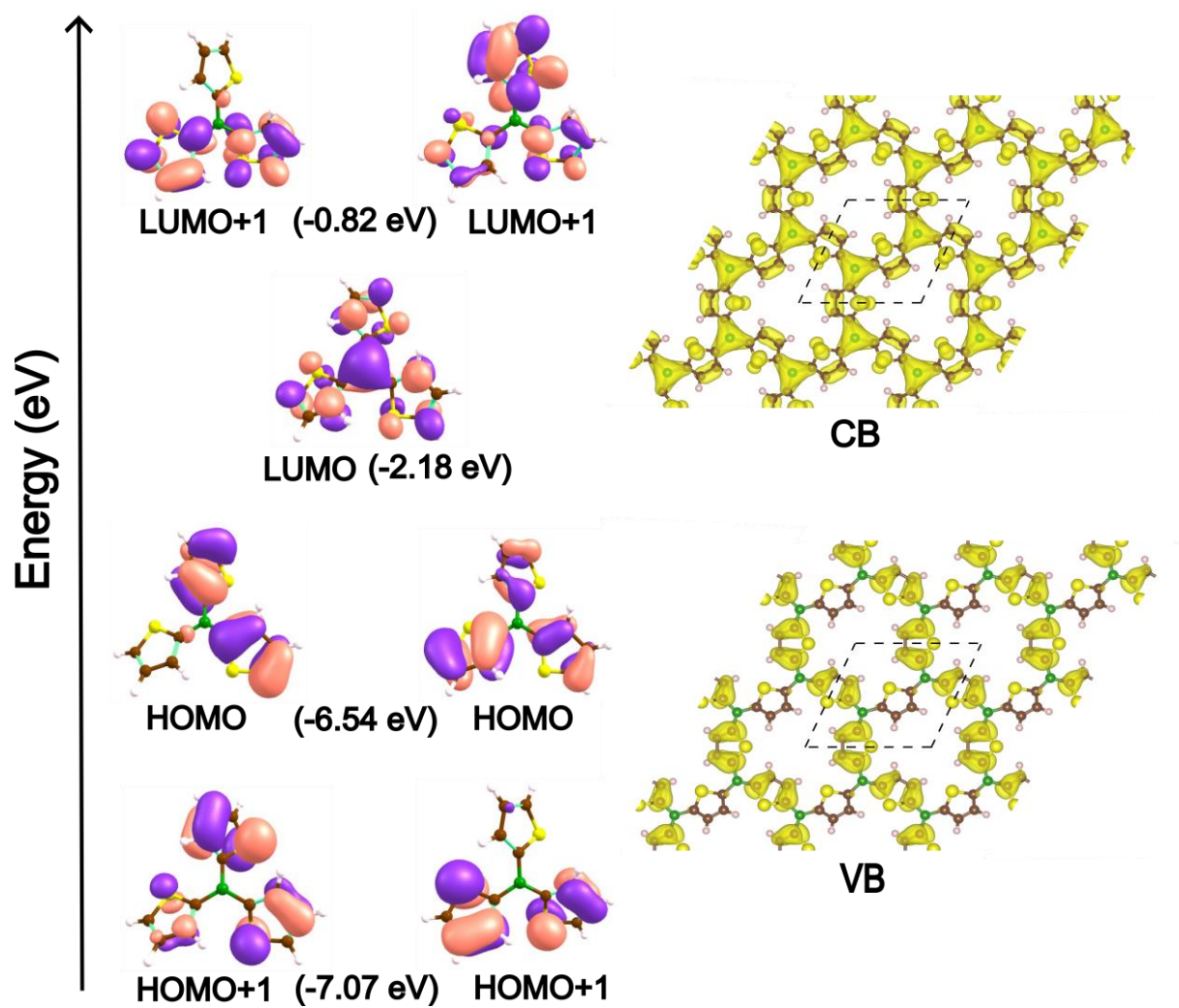

**Figure S7:** Frontier orbitals of molecular fragment of BTP-1 polymer and partial charge distribution in the CBs and VBs; the dashed rhombus indicates the unit cell.

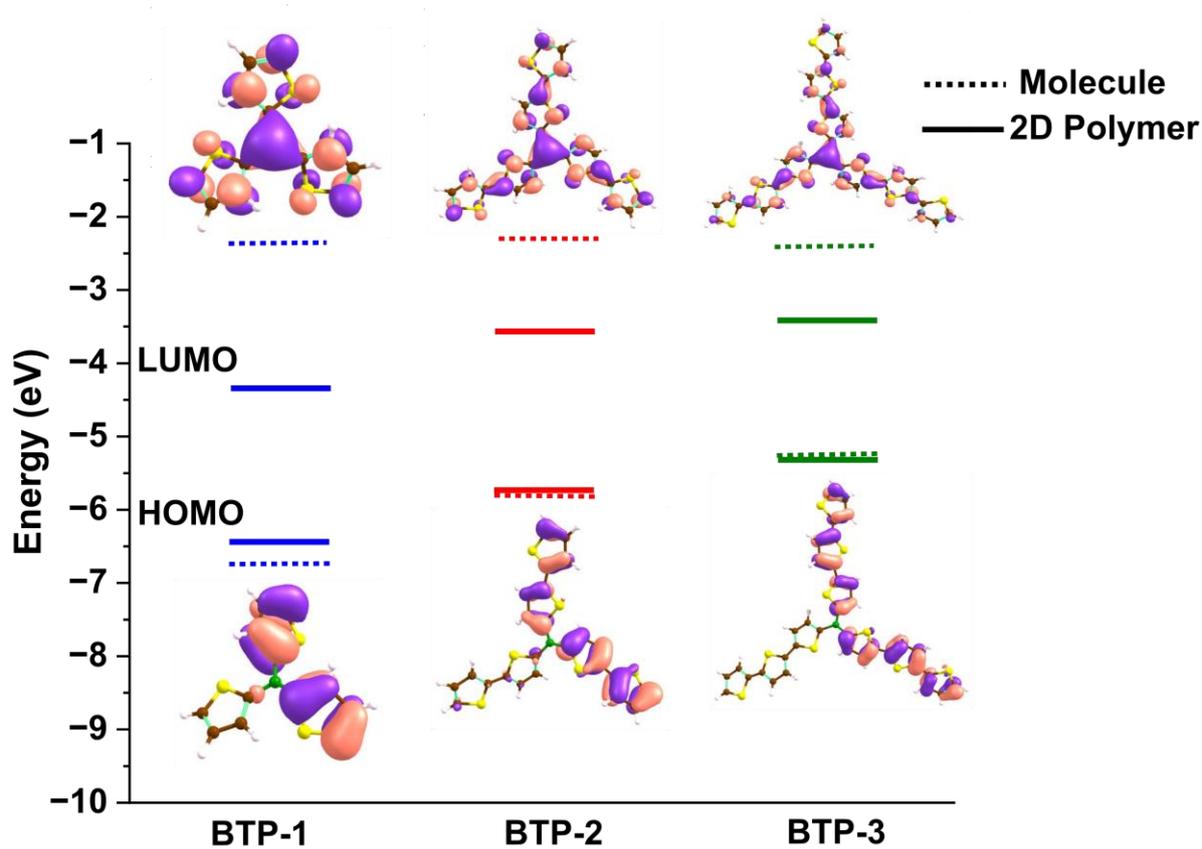

**Figure S8:** Frontier orbital energy level and band position for molecular fragments and corresponding BTP polymers. The variation in the number of thiophene units, influencing the pore size, governs the extent of π conjugation in BTP molecules. This, in turn, regulates the HOMO-LUMO gap, consequently affecting the band gap of BTP polymers. The increased π conjugation and dispersion of the frontier band contribute to a smaller band gap.

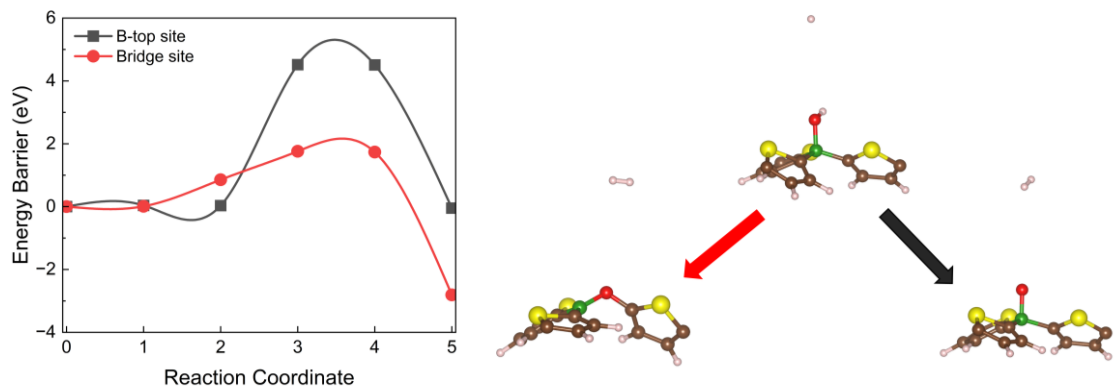

**Figure S9:** Reaction barriers form B-top *OH to B-top and bridge *O calculated using NEB method.

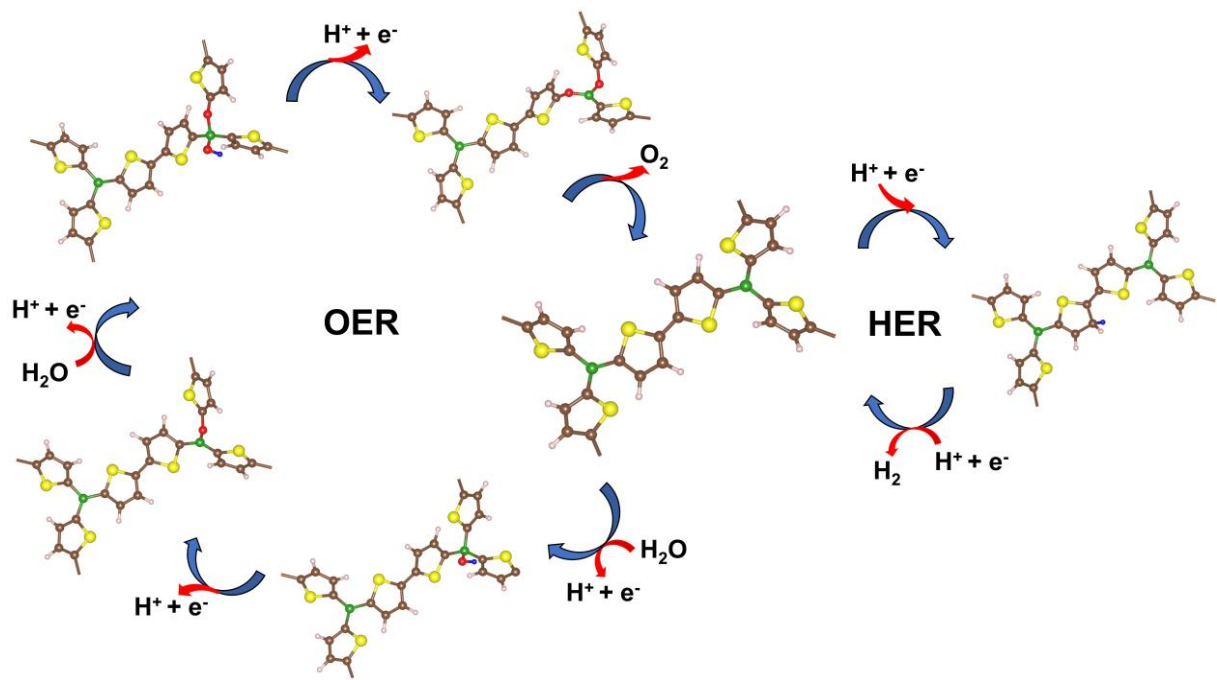

**Figure S10:** Schematics of HER and OER on BTP-2 polymer.

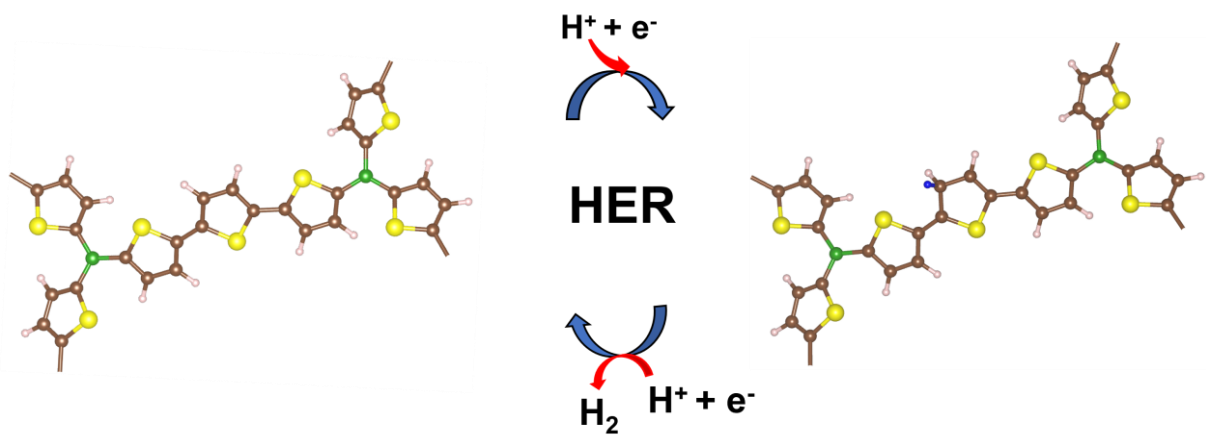

**Figure S11:** Schematics of HER on BTP-3 polymer.

**Supplementary Tables**

**Table S1:** Lattice parameters and pore size of BTP polymers.

| System | Lattice Parameters | Pore Size (Å) |
|---|---|---|
| BTP-1 | a = 9.61 = b = 9.61 Å, α = β = 90° | 4.97 |
| BTP-2 | a = 16.32 = b = 16.32 Å, α = β = 90° | 12.09 |
| BTP-3 | a = 23.17 = b = 23.17 Å, α = β = 90° | 18.36 |

**Table S2.** Free energy change of HER for BTP polymers without light irradiation.

| | ΔG (eV) | | |
|---|---|---|---|
| | BTP-1 | BTP-2 | BTP-3 |
| $* + H^+ + e^- = H^*$ | 0.48 | 0.52 | 0.57 |

**Table S3.** Free energy change of OER for BTP polymers without light irradiation.

| | ΔG (eV) | |
|---|---|---|
| | BTP-1 | BTP-2 |
| $* + H_2O = OH^* + H^+ + e^-$ | 1.60 | 1.70 |
| $OH^* = O^* + H^+ + e^-$ | 0.16 | 0.13 |
| $O^* + H_2O = O^*OH^* + H^+ + e^-$ | 1.99 | 2.02 |
| $O^*OH^* = O^*O^* + H^+ + e^-$ | 0.52 | 0.39 |
| $O^*O^* = * + O_2$ | 0.66 | 0.68 |